\documentclass{ckm}                 

\usepackage{txfonts}            

\confname{Workshop on the CKM Unitarity Triangle, IPPP Durham, April
  2003}

\title{Signals of New Physics using angular analysis in $B\to V_1 V_2$
  decays\thanks{Talk presented by Rahul Sinha.}
}

\author{David London\addressmark{a}, Nita Sinha\addressmark{b} and Rahul Sinha\addressmark{b}}

\address[a]{ Laboratoire Ren\'e J.-A. L\'evesque,
Universit\'e de Montr\'eal,
C.P. 6128, succ. centre-ville, Montr\'eal, QC,
Canada H3C 3J7}
\address[b]{Institute of Mathematical Sciences, Taramani,
 Chennai 600113, India}

\def\beq{\begin{equation}}
\def\eeq{\end{equation}}
\def\bea{\begin{eqnarray}}
\def\eea{\end{eqnarray}}
\def\nn{\nonumber}
\def\sss{\scriptscriptstyle}

\def\bd{B_d^0}
\def\bdbar{{\overline{B_d^0}}}

\def\barp{{\raise.35ex\hbox
{${\sss (}$}}~--~{\raise.35ex\hbox{${\sss )}$}}}
\def\bdbarp{\hbox{$B_d$\kern-1.4em\raise1.4ex\hbox{\barp}}}
\def\bsbarp{\hbox{$B_s$\kern-1.4em\raise1.4ex\hbox{\barp}}}
\def\barpd{{\raise.35ex\hbox
{${\sss (}$}}--{\raise.35ex\hbox{${\sss )}$}}}
\def\dbarp{\hbox{$D^{*0}$\kern-1.6em\raise1.5ex\hbox{\barpd}}}
\def\kbarp{\hbox{$K^{*0}$\kern-1.6em\raise1.5ex\hbox{\barpd}}}
\def\dpbarp{\hbox{$D^{0}$\kern-1.2em\raise1.5ex\hbox{\barpd}}}
\def\ks{K_{\sss S}}

\def\roughly#1{\mathrel{\raise.3ex\hbox
{$#1$\kern-.75em\lower1ex\hbox{$\sim$}}}}

\def\adir00{{a_{\sss dir}^{00}}}

\def\B00{B^{00}}
\def\Bp0{B^{+0}}

\def\epjc#1#2#3{{\it Eur.\ Phys.\ J.}\ {\bf C#1}, #3 (19#2)}

\def\plb#1#2#3{{\it Phys.\ Lett.} {\bf #1B}, #3 (19#2)}
\def\prd#1#2#3{{\it Phys.\ Rev.} {\bf D#1}, #3 (19#2)}
\def\newprd#1#2#3{{\it Phys.\ Rev.} {\bf D#1}: #3 (19#2)}

\def\newprl#1#2#3{{ Phys.\ Rev.\ Lett.} {\bf #1}: #3 (#2)}

\def\prl#1#2#3{{\it Phys.\ Rev.\ Lett.} {\bf #1}, #3 (19#2)}

\begin{document}

\begin{abstract}
  We show that an angular analysis of $B \to V_1 V_2$ decays yields
  numerous tests for new physics in the decay amplitudes. Many of
  these new-physics observables are nonzero even if the strong phase
  differences vanish. For certain observables, neither time-dependent
  measurements nor tagging is necessary. Should a signal for new
  physics be found, one can place a lower limit on the size of the
  new-physics parameters, as well as bound its effect on the
  measurement of the $B^0$--${\bar B}^0$ mixing phase.
\end{abstract}

\maketitle


\section{Introduction}

CP violation in the $B$ system is now firmly established.  Successful
runs at both Belle and BaBar detectors have made it possible for the
weak phase $\beta$ to be measured accurately \cite{beta}.  This
measurement of one of the interior angles of the unitarity triangle
\cite{CPreview}, is primarily performed using the so-called
``gold-plated'' mode $\bd(t)\to J/\psi \ks$. Having achieved this,
attention is now being focused on measuring $\beta$ using other modes.
In the Standard model $\bd(t) \to \phi \ks$ and $\bd(t) \to \eta'
\ks$ \cite{LonSoni} also measure $\beta$ to a good approximation. If
the values of $\beta$ measured using various modes were to disagree,
it would provide an indication of New Physics (NP).  Indeed, at
present there appears to be a discrepancy between the value of $\beta$
extracted from $\bd(t)\to J/\psi \ks$ and that obtained from $\bd(t)
\to \phi \ks$ \cite{phiKs}. Should this difference remain as more data
is accumulated, it would provide an indirect evidence for a NP
amplitude in $B \to \phi K$. It is therefore important to explore
other signals of NP, in order to corroborate this result.

Signals of NP obtained by comparing $\beta$ extracted from $\bd(t)\to
J/\psi \ks$ and $\bd(t) \to \phi \ks$, rely on the fact that the decay
amplitude for $\bd(t)\to J/\psi \ks$ is dominated by a single
contribution.  In this case, the weak-phase information can be
extracted cleanly, i.e.\ with no hadronic uncertainties \cite{foot1}.
However, this clean extraction is subject, to the absence of NP.  If
NP affects $\bd$--$\bdbar$ mixing only, the analysis is unchanged,
except that the measured value of $\beta$ is not the true SM value,
but rather one that has been shifted by a new-physics phase.  On the
other hand, if the NP affects the decay amplitude \cite{GrossWorah},
then the extraction of $\beta$ is no longer clean -- it may be
contaminated by hadronic uncertainties.

NP can affect the decay amplitude either at loop level (i.e.\
in the $b\to s$ penguin amplitude) or at tree level. Examples of such
new-physics models include non-minimal supersymmetric models and
models with $Z$-mediated flavor-changing neutral currents
\cite{newphysics}. In all cases, if the new contributions have a
different weak phase than that of the SM amplitude, then the measured
value of $\beta$, $\beta^{meas}$, no longer corresponds to the phase
of $\bd$--$\bdbar$ mixing, $\beta^{mix}$. (Note that $\beta^{mix}$
could include NP contributions to the mixing.)

If NP is present and contributes to the decay amplitude, it would be
preferable to have {\it direct} evidence for this second amplitude.
One would also like to obtain information about it (magnitude, weak
and strong phases). It is therefore important to have as many
independent tests as possible for NP.  One possibility is to search
for direct CP violation. However, direct CP asymmetries vanish if the
strong phase difference between the SM and NP amplitudes is zero.  It
has been argued that this may well be the case in $B$ decays, due to
the fact that the $b$-quark is rather heavy.  We show, however, that
if one considers $B$-meson decays to two vector mesons, $B\to V_1
V_2$, many signals for NP emerge, including several that are nonzero even if
the strong phase differences vanish.  Furthermore, if {\it any}
NP signal is found, one can place a lower bound on the size of the NP
amplitude, and on the difference $|\beta^{meas} - \beta^{mix}|$.  An
angular analysis of any of the modes such as $\bd(t) \to J/\psi K^*$
or $\phi K^*$, $D^* D_s^*$ can be used for such a study. A similar
analysis can be used within the SM to analyze decays such as $\bd(t)
\to D^{*+} D^{*-}$.

In section \ref{sec2} of the talk we examine how the large number of
observables that $B\to V_1 V_2$, decay modes provide are modified in
the presence of NP. In section \ref{sec3} we derive `12' relations,
the violation of any of which would signal NP. In section \ref{sec4}
we briefly discuss constraints on the size of NP as well as on
$|\beta^{meas} - \beta^{mix}|$, which can be obtained if NP is
observed.

\section{Observables in $\mathbf{B\to V_1 V_2}$ \label{sec2}}

Consider the decay $B\to V_1 V_2$, to which a single weak decay
amplitude contributes within the SM.  Suppose that there is a
new-physics amplitude, with a different weak phase, that contributes
to the decay. The decay amplitude for each of the three possible
helicity states may be generally written as
\begin{eqnarray}
A_\lambda \equiv Amp (B \to V_1V_2)_\lambda &=& a_\lambda e^{i
\delta_\lambda^a} + b_\lambda e^{i\phi} e^{i \delta_\lambda^b} ~,
\nn\\
{\bar A}_\lambda \equiv Amp ({\bar B} \to
{\overline{V}}_1 {\overline{V}}_2)_\lambda &=& a_\lambda e^{i
\delta_\lambda^a} + b_\lambda e^{-i\phi} e^{i \delta_\lambda^b} ~,
\label{amps}
\end{eqnarray}
where $a_\lambda$ and $b_\lambda$ represent the SM and NP amplitudes,
respectively, $\phi$ is the new-physics weak phase, the
$\delta_\lambda^{a,b}$ are the strong phases, and the helicity index
$\lambda$ takes the values $\left\{ 0,\|,\perp \right\}$. Using CPT
invariance, the full decay amplitudes can be written as
\begin{eqnarray}
{\cal A} &=& Amp (B\to V_1V_2) = A_0 g_0 + A_\| g_\| + i \, A_\perp
g_\perp~, \nn\\
{\bar{\cal A}} &=& Amp ({\bar B} \to {\overline{V}}_1
{\overline{V}}_2) = {\bar A}_0 g_0 + {\bar A}_\| g_\| - i \, {\bar
A}_\perp g_\perp~,
\label{fullamps}
\end{eqnarray}
where the $g_\lambda$ are the coefficients of the helicity amplitudes
written in the linear polarization basis. The $g_\lambda$ depend only
on the angles describing the kinematics \cite{glambda}. The above
equations enable us to write the time-dependent decay rates as
{\setlength\arraycolsep{1pt}
\begin{eqnarray}
\label{decayrates}
\Gamma(\bdbarp(t) \to V_1V_2) & = & e^{-\Gamma t}
\sum_{\lambda\leq\sigma}\; \Bigl(\Lambda_{\lambda\sigma} \pm
\Sigma_{\lambda\sigma}\cos(\Delta M t)\nn\\
&&\qquad \mp \rho_{\lambda\sigma}\sin(\Delta M t)\Bigr)\, g_\lambda
g_\sigma ~.%
\end{eqnarray}}
Thus, by performing a time-dependent angular analysis of the decay
$\bd(t) \to V_1V_2$, one can measure 18 observables. These are:
{\setlength\arraycolsep{2pt}
\begin{equation}
\begin{array}{lll}
&\Lambda_{\lambda\lambda}=\displaystyle
\frac{1}{2}(|A_\lambda|^2+|{\bar A}_\lambda|^2),&
\Sigma_{\lambda\lambda}=\displaystyle
\frac{1}{2}(|A_\lambda|^2-|{\bar A}_\lambda|^2),\nn \\[1.ex]
&\Lambda_{\perp i}= -\!{\rm Im}({ A}_\perp { A}_i^* \!-\! {\bar
A}_\perp {{\bar A}_i}^* ),
&\Lambda_{\| 0}= {\rm Re}(A_\| A_0^*\! +\! {\bar A}_\| {{\bar A}_0}^*
), \nn \\[1.ex]
&\Sigma_{\perp i}= -\!{\rm Im}(A_\perp A_i^*\! +\! {\bar A}_\perp
{{\bar A}_i}^* ),
&\Sigma_{\| 0}= {\rm Re}(A_\| A_0^*\!-\! {\bar A}_\| {{\bar A}_0}^*
),\nn\\[1.ex]
&\rho_{\perp i}\!=\! {\rm Re}\!\Bigl(\frac{q}{p} \!\bigl[A_\perp^*
{\bar A}_i\! +\! A_i^* {\bar A}_\perp\bigr]\Bigr),
&\rho_{\perp \perp}\!=\! {\rm Im}\Bigl(\frac{q}{p}\, A_\perp^*
{\bar A}_\perp\Bigr),\nn\\[1.ex]
&\rho_{\| 0}\!=\! -{\rm Im}\!\Bigl(\frac{q}{p}[A_\|^* {\bar A}_0\! +
\!A_0^* {\bar A}_\| ]\Bigr),
&\rho_{ii}\!=\! -{\rm Im}\!\Bigl(\frac{q}{p} A_i^* {\bar A}_i\Bigr),
\end{array}
\vspace*{-0.2in}
  \label{eq:obs}
\end{equation}
}
where $i=\{0,\|\}$. In the above, $q/p = \exp({-2\,i\beta^{mix}})$,
where $\beta^{mix}$ is the weak phase describing $\bd$--$\bdbar$
mixing. Note that $\beta^{mix}$ may include NP effects (in the SM,
$\beta^{mix} = \beta$). Note also that the signs of the various $\rho$
terms depend on the CP-parity of the various helicity states. We have
chosen the sign of $\rho_{00}$ and $\rho_{\|\|}$ to be $-1$, which
corresponds to the final state $J/\psi K^*$.

The 18 observables given above can be written in terms of 13
theoretical parameters: three $a_\lambda$'s, three $b_\lambda$'s,
$\beta^{mix}$, $\phi$, and five strong phase differences defined by
$\delta_\lambda \equiv \delta_\lambda^b - \delta_\lambda^a$, $\Delta_i
\equiv \delta_\perp^a - \delta_i^a$. The explicit expressions for the
observables are as follows:
{
{\setlength\arraycolsep{1pt}
\begin{eqnarray}
\Lambda_{\lambda\lambda} &&= a_\lambda^2 + b_\lambda^2 + 2 a_\lambda
b_\lambda \cos\delta_\lambda \cos\phi ~,\nn \\
\Sigma_{\lambda\lambda}  &&= - 2 a_\lambda b_\lambda
\sin\delta_\lambda \sin\phi ~, \nn\\
\Lambda_{\perp i} &&= 2 \left[ a_\perp b_i \cos(\Delta_i - \delta_i)
- a_i b_\perp \cos(\Delta_i + \delta_\perp) \right] \sin\phi ~,\nn\\
\Lambda_{\| 0} &&= 2 \big[ a_\| a_0 \cos(\Delta_0 - \Delta_\|)+
   a_\| b_0 \cos(\Delta_0 - \Delta_\| - \delta_0) \cos\phi  \nn\\
&&~~~~~ + a_0 b_\| \cos(\Delta_0 - \Delta_\| + \delta_\|)\cos\phi \nn\\
&&~~~~~+ b_\| b_0 \cos(\Delta_0 - \Delta_\| + \delta_\| - \delta_0)\big] ~,\nn\\
\Sigma_{\perp i} &&= -2 \big[ a_\perp a_i \sin \Delta_i + a_\perp
   b_i \sin(\Delta_i - \delta_i) \cos\phi \nn \\
&&~+ a_i b_\perp \sin(\Delta_i + \delta_\perp) \cos\phi +
   b_\perp b_i \sin (\Delta_i + \delta_\perp - \delta_i) \big] ~, \nn\\
\Sigma_{\| 0} &&= 2 \big[ a_\| b_0 \sin(\Delta_0 - \Delta_\| -\delta_0) \nn\\
&&~~~~~- a_0 b_\| \sin(\Delta_0 - \Delta_\| + \delta_\|) \big]\sin\phi ~, \nn\\
\rho_{ii} &&= a_i^2 \sin 2\beta^{mix} + b_i^2 \sin(2\beta^{mix} + 2 \phi) \nn\\
&&~~~~~ + 2 a_i b_i \cos\delta_i \sin(2\beta^{mix} + \phi)~, \nn \\
\rho_{\perp\perp} &&= - a_\perp^2 \sin 2\beta^{mix}
 - b_\perp^2 \sin(2\beta^{mix} + 2\phi)\nn \\
&&~~~~~ - 2 a_\perp b_\perp\cos\delta_\perp \sin(2 \beta^{mix} + \phi)~, \nn\\
\rho_{\perp i} &&= 2 \big[ a_i a_\perp \cos \Delta_i \cos 2\beta^{mix}\nn\\
&&~~~~~+a_\perp b_i \cos(\Delta_i - \delta_i) \cos(2 \beta^{mix} + \phi) \nn \\
&&~~~~~+ a_i b_\perp \cos(\Delta_i + \delta_\perp) \cos(2 \beta^{mix}+\phi) \nn\\
&&~~~~~+ b_i b_\perp \cos(\Delta_i + \delta_\perp - \delta_i)
     \cos(2\beta^{mix} + 2\phi) \big] ~,\nn\\
\rho_{\| 0} &&= 2 \big[ a_0 a_\| \cos(\Delta_0 - \Delta_\|)
     \sin 2\beta^{mix}\nn\\
&&~~~~~+ a_\| b_0 \cos(\Delta_0 -\Delta_\| - \delta_0)
     \sin(2\beta^{mix}+\phi) \nn\\
&&~~~~~+ a_0 b_\| \cos(\Delta_0 - \Delta_\| + \delta_\|)
     \sin(2\beta^{mix} + \phi) \nn\\
&&+ b_0 b_\| \cos(\Delta_0 - \Delta_\| + \delta_\| -\delta_0)
     \sin(2\beta^{mix} + 2\phi) \big] ~.
\label{observables}
\end{eqnarray} }
}

It is straightforward to show that one cannot extract $\beta^{mix}$.
There are a total of six amplitudes describing $B \to V_1 V_2$ and
${\bar B} \to {\overline{V}}_1 {\overline{V}}_2)$ decays
[Eq.~(\ref{amps})]. Thus, at best one can measure the magnitudes and
relative phases of these six amplitudes, giving 11 measurements. Since
the number of measurements (11) is fewer than the number of theoretical
parameters (13), one cannot obtain any of the theoretical unknowns
purely in terms of observables. In particular, it is impossible to
extract $\beta^{mix}$ cleanly.

\section{Signals of New Physics \label{sec3}}

In the absence of NP, $b_\lambda = 0$. The number of parameters is
then reduced from 13 to 6: three $a_\lambda$'s, two strong phase
differences ($\Delta_i$), and $\beta^{mix}$. All of these can be
determined cleanly in terms of observables. There are 18 observables,
but only 6 theoretical parameters, thus 12 relations must exist among
the observables in the absence of NP. (Of course, only five of these
are independent.) These 12 relations are:
\begin{eqnarray}
&& \Sigma_{\lambda\lambda}= \Lambda_{\perp i}= \Sigma_{\|
  0}=0 \nn\\
&& \frac{\rho_{ii}}{\Lambda_{ii}} =
-\frac{\rho_{\perp\perp}}{\Lambda_{\perp\perp}} =
\frac{\rho_{\|0}}{\Lambda_{\| 0}}\nn \\
&&
\Lambda_{\|0}=\frac{1}{2\Lambda_{\perp\perp}}\Bigl[
  \frac{\Lambda_{\lambda\lambda}^2\rho_{\perp 0}
  \rho_{\perp\|}+\Sigma_{\perp 0}\Sigma_{\perp
  \|}(\Lambda_{\lambda\lambda}^2
  -\rho_{\lambda\lambda}^2)}
 {\Lambda_{\lambda\lambda}^2-\rho_{\lambda\lambda}^2}\Bigr]\nn
  \\
&&
  \frac{\rho_{\perp i}^2}{4\Lambda_{\perp\perp}\Lambda_{i
      i}-\Sigma_{\perp i}^2}=\frac{\Lambda_{\perp\perp}^2
    -\rho_{\perp\perp}^2}{\Lambda_{\perp\perp}^2}~.
  \label{eq:no_np}
\end{eqnarray}
The important consequence is \cite{London:2003rk} that {\em the
  violation of any of the above relations will be a smoking-gun signal
  of NP.}  It may be emphasized that the angular analysis of $B\to V_1
V_2$ decays provides numerous tests for the presence of NP.

The observable $\Lambda_{\perp i}$ deserves special attention
\cite{DattaLondon}. From Eq.~(\ref{observables}), one sees that even
if the strong phase differences vanish, $\Lambda_{\perp i}$ is nonzero
in the presence of NP ($\phi\ne 0$), in stark contrast to the
direct CP asymmetries (proportional to $\Sigma_{\lambda\lambda}$).
This is due to the fact that the $\perp$ helicity is CP-odd, while the
$0$ and $\|$ helicities are CP-even.  While the reconstruction
of the full $\bd(t)$ and $\bdbar(t)$ decay rates in
Eq.~(\ref{decayrates}) requires both tagging and time-dependent
measurements, the $\Lambda_{\lambda\sigma}$ terms survive even if the
two rates for $\bd(t)$ and $\bdbar(t)$ decays are added together. We
note also that these terms are time-independent.  Therefore, {\it no
  tagging or time-dependent measurements are needed to extract
  $\Lambda_{\perp i}$}! It is only necessary to perform an angular
analysis of the final state $V_1 V_2$. Thus, this measurement can even
be made at a symmetric $B$-factory.  The decays of charged $B$ mesons
to vector-vector final states are even simpler to analyze since no
mixing is involved.  One can in principle combine charged and neutral
$B$ decays to increase the sensitivity to NP. A nonzero value
of $\Lambda_{\perp i}$ would provide a clear signal for NP
\cite{FPCP}.

The decays of both charged and neutral $B$ mesons to $D^* D_s^*$ can
be analyzed similarly. Because these modes are dominated by a single
decay amplitude in the SM, no direct CP violation is expected. Further,
since this is not a final state to which both $\bd$ and $\bdbar$ can
decay, no indirect CP violation is possible either. An angular
analysis of these decays would therefore be very interesting in
exploring the presence of NP.

It must be noted that, despite the large number of new-physics
signals, it is still possible for the NP to remain hidden. This
happens if a singular situation is realized. If the three strong phase
differences $\delta_\lambda$ vanish, and the ratio $r_\lambda \equiv
b_\lambda/a_\lambda$ is the same for all helicities, i.e. $r_0 = r_\|
= r_\perp$, then it is easy to show that the relations in
Eq.~(\ref{eq:no_np}) are all satisfied. Thus, if the NP happens to
respect these very special conditions, the angular analysis of $B\to
V_1 V_2$ would show no signal for NP, yet the measured value of
$\beta$ would not correspond to the actual phase of $\bd$--$\bdbar$
mixing.

\section{Constraints on the size of New Physics \label{sec4}}

We have argued earlier, that in the presence of NP one cannot extract
the true value of $\beta^{mix}$. However, as we will describe below,
the angular analysis does allow one to constrain the value of the
difference $|\beta^{\mathit{meas}} - \beta^{mix}|$, as well as the
size of the NP amplitudes $b^2_\lambda$. Naively, it appears
impossible to obtain any constraints on the NP parameters, since we
have 11 measurements, but 13 theoretical unknown parameters.  However,
because the equations are nonlinear, such constraints are possible.
Below, we list some of these constraints \cite{London:2003rk}

In the constraints, we will make use of the following quantities. For
the vector-vector final state, the analogue of the usual direct CP
asymmetry is $a_{\lambda}^{dir} \equiv
\Sigma_{\lambda\lambda}/\Lambda_{\lambda\lambda}$, which is
helicity-dependent. For convenience, we define the related quantity
$y_\lambda^2 = (1 - \Sigma_{\lambda\lambda}^2/
  \Lambda_{\lambda\lambda}^2)$. The measured value of $\sin 2\beta$
can also depend on the helicity of the final state:
$\rho_{\lambda\lambda}$ can be recast in terms of a measured weak
phase $2\beta^{\mathit{meas}}_{\lambda}$, defined as
$\sin2\,\beta^{\mathit{meas}}_{\lambda}=\pm
\rho_{\lambda\lambda}/(\Lambda_{\lambda\lambda} y_\lambda)$,
%
%
where the $+$ $(-)$ sign corresponds to $\lambda=0,\|$ ($ \perp$). In
terms of these quantities, the size of NP amplitudes $b_\lambda^2$ may
be expressed as
\beq
2\,b_\lambda^2\,\sin^2\phi = \Lambda_{\lambda\lambda}
    \Big(1-y_\lambda\cos(2\beta^{meas}_\lambda-2\beta)\Big) ~.
\label{eq:a-b-2}
\eeq

The form of the constraints depends on which new-physics signals are
observed; we give a partial list below. For example, suppose that
direct CP violation is observed in a particular helicity state. In
this case a lower bound on the corresponding NP amplitude can be
obtained by minimizing $b^2_\lambda$ with respect to $\beta$ and
$\phi$:
\beq
b^2_\lambda \ge {1\over 2} \Lambda_{\lambda\lambda} \left[ 1 -
y_\lambda \right].
\eeq
On the other hand, suppose that the new-physics signal is
$\beta^{\mathit{meas}}_i \ne \beta^{\mathit{meas}}_j$. Defining
$2\omega \equiv 2\beta^{\mathit meas}_j-2\beta^{\mathit meas}_i$ and
$\eta_\lambda \equiv 2 ( \beta^{\mathit{meas}}_\lambda - \beta^{mix}
)$, the minimization of $(b_i^2 \mp b_j^2)$ with respect to $\eta_i$
and $\phi$ yields
\beq
(b_i^2 \mp b_j^2) \ge \frac{\Lambda_{ii} \mp \Lambda_{jj}}{2} - \frac{
\left\vert y_i \Lambda_{ii} \mp y_j \Lambda_{jj} e^{2 i \omega}
\right\vert }{2} ~,
\label{eq:bsq-omega-bounds}
\eeq
where $\Lambda_{ii} > \Lambda_{jj}$ is assumed. If there is no direct
CP violation ($\Sigma_{\lambda\lambda} = 0$), but $\Lambda_{\perp i}$
is nonzero, one has
\beq
2 (b_i^2 \mp b_\perp^2) \geq \Lambda_{ii} \mp \Lambda_{\perp\perp} -
\sqrt{ \left( \Lambda_{ii} \mp \Lambda_{\perp\perp}\right)^2 \pm
\Lambda_{\perp i}^2} ~.
\label{eq:bsq-Lambda-bounds}
\eeq
%

One can also obtain bounds on $|\beta^{\mathit{meas}}_\lambda -
\beta^{mix}|$, though this requires the nonzero measurement of
observables involving the interference of different helicities. For
example, if $\Lambda_{\perp i}$ is nonzero and
$\Sigma_{\lambda\lambda} = 0$, we find
\[
\Lambda_{ii} \cos\eta_i + \Lambda_{\perp\perp} \cos(\eta_\perp - 2
\eta_i) \le \sqrt{ \left( \Lambda_{ii} + \Lambda_{\perp\perp}
\right)^2 - \Lambda_{\perp i}^2}~,
\]
\null\vskip-12truemm
\beq
\Lambda_{ii} \cos\eta_i - \Lambda_{\perp\perp} \cos \eta_\perp
\le \sqrt{ \left( \Lambda_{ii} - \Lambda_{\perp\perp}
\right)^2 + \Lambda_{\perp i}^2}~.
\eeq
If $\Lambda_{\perp i} \ne 0$, one cannot have $\eta_i = \eta_\perp =
0$. These constraints therefore place a lower bound on
$|\beta^{\mathit{meas}}_i - \beta^{mix}|$ and/or
$|\beta^{\mathit{meas}}_\perp - \beta^{mix}|$.

A-priori, one does not know which of the above constraints is the
strongest -- this depends on the actual values of the observables. Of
course, in practice, one will simply perform a fit to obtain the best
lower bounds on these NP parameters \cite{London:2003rk}. However, it is
interesting to study analytically the dependence of constraints as a
function of observables which would signal NP if non-zero.

If the apparent discrepancy in the value of $\sin 2\beta$ as obtained
from measurements of $\bd(t)\to J/\psi \ks$ and $\bd(t) \to \phi \ks$
\cite{phiKs} is on account of NP, angular analyses of $\bd(t)\to
J/\psi K^*$ and $\bd(t) \to \phi K^*$ would allow one to determine if
NP is indeed present. If NP signal is confirmed, this analysis would
allow one to put constraints on the NP parameters.

Finally, we note that this analysis can also be applied within the SM
to decays such as $\bd(t) \to D^{*+} D^{*-}$. These decays have both a
tree and a penguin contribution, so that $\beta^{mix}$ cannot be
extracted cleanly. Assuming no NP, the above analysis allows
one to obtain lower bounds on the ratio of penguin to tree amplitudes,
as well as on $|\beta^{\mathit{meas}}_\lambda - \beta^{mix}|$. This
can serve as a cross-check on other measurements of $\beta^{mix}$, as
well as on model calculations of the hadronic amplitudes.

\section{Summary \label{sec5}}

In the standard model (SM), the cleanest extraction of the CP angles
comes from neutral $B$ decays that are dominated by a single decay
amplitude. If there happens to be a new-physics (NP) contribution to
the decay amplitude, with a different weak phase, this could seriously
affect the cleanliness of the measurement. There is already a hint of
such NP, as indicated by the discrepancy between the value of $\beta$
extracted from $\bd(t)\to J/\psi \ks$ and that obtained from $\bd(t)
\to \phi \ks$. However, it is important to confirm this through
independent direct tests, and to make an attempt to obtain information
about the NP amplitude, if possible.

We have shown that this type of NP can be probed by
performing an angular analysis of the related $B \to V_1 V_2$ decay
modes. There are numerous relations that are violated in the presence
of NP, and several of these signals remain nonzero even if the strong
phase difference between the SM and NP amplitudes vanishes. The most
incisive test is a measurement of $\Lambda_{\perp i} \ne 0$. To obtain
this observable, neither tagging nor time-dependent measurements are
necessary -- one can combine all neutral and charged $B$ decays.

Furthermore, should a signal for NP be found, one can place a
lower bound on the difference $|\beta^{\mathit{meas}} - \beta^{mix}|$,
as well as on the size of the NP amplitudes. By applying this analysis
to the decays $\bd(t)\to J/\psi K^*$ and $\bd(t) \to \phi K^*$, one
can confirm the presence of the NP that is hinted at in the
measurements of $\bd(t)\to J/\psi \ks$ and $\bd(t) \to \phi \ks$
\cite{phiKs}. It can even be applied within the SM to analyze decays
such as $\bd(t) \to D^{*+} D^{*-}$, which receive both tree and
penguin contributions.

N.S. and R.S. would like to thank Prof. Patricia Ball and the local
organizers of the CKM Workshop for financial support. We thank the
organizers for an exciting conference. The work of D.L. was
financially supported by NSERC of Canada. The work of N.S. was
supported by the Department of Science and Technology, India.


\begin{thebibliography}{9}
\bibitem{foot1} In fact, there are two weak amplitudes that can
  contribute even to $\bd(t)\to J/\psi \ks$: the tree amplitude and the
  $b\to s$ penguin amplitude. However, the weak phases of these two
  amplitudes are equal (they vanish in the Wolfenstein parametrization
  of the CKM matrix \cite{Wolfenstein}), so that there is effectively
  only a single weak amplitude contributing to $\bd\to J/\psi
  \ks$. Thus, the extraction of the CP phase $\beta$ from this decay
  mode is extremely clean.
\end{thebibliography}
\end{document}